\documentstyle[prl,aps]{revtex}
\draft
\begin{document}

\title{Screening effects in superconductors}
\author{M. Capezzali$^{(a)}$, D. Ariosa$^{(b)}$ and H.Beck$^{(a)}$}
\address{$^{(a)}$ Institut de Physique, Universit\'{e} de Neuch\^{a}tel, Rue A.L.
Breguet 1,
2000 Neuch\^{a}tel, Switzerland \\
$^{(b)}$ IBM Research Division, S\"{a}umerstr. 4, 8803 R\"{u}schlikon (ZH), 
Switzerland}
\maketitle
\widetext
\begin{abstract}
\begin{center}
\parbox{14cm}{The partition function of the Hubbard model with local attraction 
and long range Coulomb repulsion between electrons is written as a functional integral 
with an action $A$ involving a pairing field $\Delta$ and a local potential $V$. 
After integration over $V$ and over fluctuations in $|\Delta|^{2}$, 
the final form of $A$ involves a Josephson coupling between the local 
phases of  $\Delta$ and a "kinetic energy" term, representing 
the screened Coulomb interaction between charge fluctuations. 
The competition between Josephson coupling and charging energy 
allows to understand the relation between $T_{C}$ and composition in high-$T_{C}$ 
materials, in particular superlattices, alloys and bulk systems of low doping.
}
\end{center}
\end{abstract}

\vspace*{1.0truecm}

We start from a Hamiltonian describing charge carriers (electrons or holes) on a lattice, 
subject to an on-site attraction, $- U$, and the long range Coulomb repulsion,
$e^{2}V_{C}$, acting on particles on different sites :
\begin{eqnarray}
H=H_{0}+H_{U}+H_{C}
\end{eqnarray}
\begin{eqnarray}
H_{0}=\sum_{\left\langle l,l' \right\rangle}^{}{
\sum_{\sigma}^{}{tc_{l,\sigma}^{+}c_{l',\sigma}^{}}}-\mu\sum_{l}^{}{
N_{l}}
\end{eqnarray}
\begin{eqnarray}
H_{U}=-{U \over 2}\sum_{l}^{}{\sum_{\sigma}^{}{
c_{l,\sigma}^{+}c_{l,\sigma}^{}c_{l,-\sigma}^{+}c_{l,-\sigma}^{}}}
\end{eqnarray}
\begin{eqnarray}
H_{C}={1 \over 2}\sum_{l\neq l'}^{}{
(N_{l}-n_{0})e^{2}V_{C}(l,l')(N_{l'}-n_{0})}.
\end{eqnarray}
Here $c_{l,\sigma}^{+}$ ($c_{l,\sigma}^{}$) are creation (annihilation) operators 
for charge carriers with spin $\sigma$ at lattice site $l$,
$\left\langle l,l' \right\rangle$ denotes pairs of nearest neighbor sites,
$\mu$ is the chemical potential, $n_{0}$ the background, neutralizing 
the density of charge carriers, and 
$N_{l}=\sum_{\sigma}^{}{c_{l,\sigma}^{+}c_{l,\sigma}^{}}$. 
The partition function can be written as a functional integral by 
means of two successive Stratonovich-Hubbard transformations
\cite{kopec,ambe}, decoupling the two
interaction terms in $H$ with the help of a complex field $\Delta$ and a real field $V$ :
\begin{eqnarray}
Z=Tr\left\{ e^{-\beta H} \right\}=\int{D^{2}\Delta\int{DV Tr\left\{ 
e^{-\beta H_{0}}T_{\tau}e^{-i\int_{0}^{-i\beta}{
d\tau\left[ \tilde{H}(\Delta,V,\tau)+\epsilon(\Delta,V,\tau) \right]}} \right\}}},
\end{eqnarray}
\begin{eqnarray}
\tilde{H}(\Delta,V,\tau)=\sum_{l}^{}{\left[ \Delta^{*}(l,\tau)c^{}_{l,\uparrow}c^{}_{l,\downarrow}
+\Delta(l,\tau)c^{+}_{l,\downarrow}c^{+}_{l,\uparrow}+iV(l,\tau)
\left(N_{l}(\tau)-n_{0} \right) \right]
}
\end{eqnarray}
\begin{eqnarray}
\epsilon(\Delta,V,\tau)={1\over U}\sum_{l}^{}{
|\Delta(l,\tau)|^{2}}+{1\over 2e^{2}}\sum_{l\neq l'}^{}{V(l,\tau)V_{C}^{-1}(l,l')V(l',\tau)
}.
\end{eqnarray}
We then evaluate the trace over the electronic degrees of freedom 
\begin{eqnarray}
Tr\left\{ 
e^{-\beta H_{0}}T_{\tau}e^{-i\int_{0}^{-i\beta}{
d\tau\left[ \tilde{H}(\Delta,V,\tau)+\epsilon(\Delta,V,\tau) \right]}} \right\}=
e^{-i\int_{0}^{-i\beta}{d\tau F(\Delta,V,\tau)}}
\end{eqnarray}
and we expand the "free energy" $F$ up to fourth order in $\Delta$ 
and to second order in $V$ and up to leading terms in space 
and time gradients of the two fields :
\begin{eqnarray}
F=F_{0}+F_{\Delta,V}
\end{eqnarray}
\begin{eqnarray}
F_{\Delta,V}(\tau)=\sum_{l}^{}{
\left[ a|\Delta(l,\tau)|^{2}-id\Delta^{*}(l,\tau)\left( 
{\partial \over \partial \tau}{}-2V(l,\tau) \right) \Delta(l,\tau)\right]}
+\nonumber \\
c\sum_{\left\langle l\neq l' \right\rangle}^{}{
|\Delta(l,\tau)-\Delta(l',\tau)|^{2}}+i\sum_{l}^{}{
V(l,\tau)\left( \left\langle N_{l} \right\rangle-n_{0} \right)}
+ \nonumber \\
{1\over 2e^{2}}\sum_{l,l'}^{}{
V(l,\tau)V_{SC}^{-1}(l,l')V(l',\tau)}+b\sum_{l}^{}{|\Delta(l,\tau)|^{4}}
\end{eqnarray}
Here the coefficients $a$, $b$, $c$, $d$ are related to the free electron 
particle-particle propagator \cite{kopec,ambe,drech,randeria,pedersen}, 
$F_{0}$ is the free 
electron contribution, and $V_{SC}^{-1}(l,l')=V_{C}^{-1}(l,l')+
\chi_{0}(l,l')$ is the screened Coulomb potential 
(approximated by its static limit) with $\chi_{0}(l,l')$ being the electronic polarizability.  
Integrating over the electric potential $V$ yields
\begin{eqnarray}
F_{\Delta}(\tau)=\sum_{l}^{}{\left[ a|\Delta(l,\tau)|^{2}+b|\Delta(l,\tau)|^{4}
-id\Delta^{*}(l,\tau){\partial {\Delta(l,\tau)}\over \partial \tau} \right] 
} \nonumber \\
+c\sum_{\left\langle l,l' \right\rangle}^{}{|\Delta(l,\tau)-\Delta(l',\tau)|^{2}} +
{1\over 2e^{2}}\sum_{l,l'}^{}{\rho(l,\tau)V_{SC}(l,l')\rho(l',\tau)}
\end{eqnarray}
where $\rho(l,\tau)=2d|\Delta(l,\tau)|^{2}+\left\langle N_{l} \right\rangle-n_{0}$ 
represents charge density fluctuations. Next we introduce amplitude and phase of 
$\Delta$. In the following, we assume to be in a relatively strong coupling 
regime ($t\ll U$) in which fermions bind into onsite singlet pairs at a temperature 
on the order of the mean field transition temperature $T_{mf}$, 
well above the superconducting phase transition, the latter being finally
triggered by the onset of phase order \cite{emery,doniach,roddick}. 
Below $T_{mf}$, $a<0$, 
so that the average amplitude has a non-vanishing mean value 
$\Delta_{0}$ given by $a+2b\Delta_{0}^{2}$. 
Charge neutrality implies $2d\Delta_{0}^{2}+\left\langle N_{l} \right\rangle
-n_{0}=0$ and, for $t\ll U$ \cite{kopec,drech,randeria,pedersen,balseiro} :
\begin{eqnarray}
c={2t^{2}\over U^{3}}, \nonumber \\
d={1\over U^{2}}, \nonumber \\
\Delta_{0}^{2}={n_{0}(2-n_{0})\over 4}U^{2} \approx {n_{0}\over 2}U^{2} 
\qquad for\quad n_{0}\ll 1.
\end{eqnarray}
Splitting the number of pairs at a given site into
$|\Delta(l,\tau)|^{2}=\Delta_{0}^{2}+{U^{2}\over 2}\delta n_{p}(l,\tau)$,  
we integrate over $\delta n_{p}(l,\tau)$ which (neglecting gradient terms) 
yields a free energy functional for phase fluctuations only :
\begin{eqnarray}
F_{\Theta}(\tau)=J\sum_{\left\langle l,l' \right\rangle}^{}{
\left[ 1-\cos{\left( \Theta(l,\tau)-\Theta(l',\tau) \right)}\right]}+{n_{0}\over 2}\sum_{l}^{}{
{\partial {\Theta(l,\tau)}\over \partial \tau}}-\nonumber \\
{1\over 2}\sum_{l,l'}^{}{{\partial {\Theta(l,\tau)}\over \partial \tau}
{1\over (2e)^{2}}W^{-1}(l,l'){\partial {\Theta(l',\tau)}\over \partial \tau}}
\end{eqnarray}
with the Josephson coupling \cite{kopec,balseiro} $J=2c\Delta_{0}^{2}=2n_{0}
\left( {t^{2}\over U} \right)$, and 
\begin{eqnarray}
W(l,l')=\cases{
V_{SC}(l,l') & for $l\neq l'$ \cr
(2e)^{2}{b\over d^{2}} & for $l=l'$
}.
\end{eqnarray}
Our expansion of $F$ in powers of $\Delta$ has yielded the on-site repulsion
$(2e)^{2}{b\over d^{2}}$ between pairs. However, due to the exclusion principle, 
two pairs cannot really sit on the same lattice site. 
Thus, in the following, we exclude $l = l'$ in the last term of (13). 
By going from the "phase velocities" 
${\partial {\Theta}\over \partial \tau}$ to the conjugate momenta
$p(l,\tau)$ we end up with the partition function of the Hamiltonian \cite{fisher}
\begin{eqnarray}
H={1\over 2}\sum_{l,l'}^{}{
\left( p(l)-{n_{0}\over 2} \right)\left[ (2e)^{2}W(l,l') \right]
\left( p(l')-{n_{0}\over 2} \right)}+J\sum_{\left\langle l,l' \right\rangle}^{}{
\left[ 1-\cos{\left( \Theta(l)-\Theta(l') \right)} \right]}.
\end{eqnarray}
Hamiltonian (15) describes the relevant physics of short coherence length 
superconductors in terms of Josephson coupled spatial phase variations and "charge 
fluctuations" coupled by the screened Coulomb interaction. 
It is also the "phase-only" representation of the Hamiltonian 
of interacting bosons \cite{fisher,cha,MPA2}. 
Its critical behavior, in particular the influence of the "background", $-{n_{0}\over 2}$, 
has been studied in these references. Here, we apply expression (15) 
to calculating the transition temperature of strongly anisotropic superconductors, such as superlattices and bulk systems in the underdoped regime \cite{tallon}. 
We make the following approximations : i) $H$ is restricted to one superconducting 
layer; ii) the screened Coulomb interaction, which takes into 
account the electric coupling between layers, is modelled by a Yukawa-form, 
with a Thomas-Fermi screening length $\lambda_{TF}$, 
depending on the density $n_{0}$ of charge carriers according to 
the wellknown formula $\lambda_{TF}={1\over 2\pi}{2.95\over\sqrt{r_{s}/a_{0}}}$
[\.{A}$^{-1}$], where $r_{s}=\left( {3 \over 4\pi n_{0}} \right)^{{1\over 3}}$ 
and $a_{0}$ is the Bohr radius; iii) considering only $n_{0}\ll 1$, 
the "background shift", $-{n_{0}\over 2}$, in the first term of (15) is neglected. 
We make connection with previous work \cite{beck,ariosa} 
by mapping (15) onto a "capacity model":
\begin{eqnarray}
{1\over 2}\sum_{l,l'}^{}{p(l)\left[ (2e)^{2}W(l,l') \right]p(l')}\approx 
{(2e)^{2}\over 2C}\sum_{l}^{}{p(l)^{2}},
\end{eqnarray}
with ${1\over 2C}={1\over e^{2}}\sum_{l}^{}{W(l,0)}={2\pi\lambda_{TF}\over\epsilon a_{L}^{2}}
e^{-{a_{L}\over\lambda_{TF}}}$.\\
This is the XY-model with kinetic energy \cite{beck,ariosa}, $\epsilon$ 
being the dielectric constant of the interlayer material and $a_{L}$ 
the lattice constant. In Ref. [14] the critical temperature in two dimensions was evaluated 
in the "self-consistent harmonic approximation" (SCHA) which gives results in good 
agreement with Monte Carlo simulations \cite{jose}. A good overall fit of the numerical 
SCHA result is \cite{marti} 
$T_{C}(\alpha)\approx T_{C}(0)\sqrt{1-{\alpha\over\alpha_{C}}}$ where 
$\alpha={(2e)^{2}\over 2CJ}$ 
is the ratio between charging and Josephson energy. When $\alpha$ 
approches $\alpha_{C}$=6.2, $T_{C}$ goes to zero. 
This approach has been successfully applied \cite{beck,ariosa} 
to calculating the superconducting transition temperature for superlattices 
and for alloys by determining the appropriate effective capacity through electrostatic 
considerations.\\
We finally use Hamiltonian (15) to find $T_{C}$ in function of doping for 
high-$T_{C}$ superconductors in the underdoped regime. 
For YBa$_{2}$Cu$_{3}$O$_{7}$, our "capacity model" has allowed to fit 
the $T_{C}$ variation of both, superlattices and alloys, in a coherent way, 
using SCHA \cite{beck,ariosa} for $J$=120$\dot{}$K at optimal doping 
($n_{0}\approx$ 0.16 of holes per cell \cite{tallon}). Below optimal doping, 
the ratio $\alpha$ increases when the number $n_{0}$ of charge carriers is reduced : 
the $n_{0}$-dependence of $J$ is given following Eqn. (13) and $C$ varies with 
$n_{0}$ through the screening length $\lambda_{TF}$. 
Using the above expression for $T_{C}(\alpha)$, with doping-dependent $J$ and $C$, 
we then find that $T_{C}$ should be zero for $n_{0}\approx$ 0.07, in good agreement with the measured phase diagram \cite{tallon}. This shows that the phase boundary in the underdoped regime can be understood in terms of Bose-Einstein condensation of preformed pairs, 
$T_{C}$ being suppressed by phase fluctuations, when the minimum doping is approached.\\
In summary, starting from the attractive Hubbard model with long range Coulomb repulsion, we
have given a microscopic derivation of a description of short coherence length 
superconductors in terms of the superconducting phase, 
the Hamiltonian for which includes a "charging energy" and a Josephson coupling. 
This is a microscopic justification of such a Hamiltonian, which has been used 
previously \cite{beck,ariosa} to calculating $T_{C}$ 
for superlattices and alloys and also allows understanding the phase 
boundary of bulk oxides in the underdoped regime.\\
This work was supported by the Swiss National Science Foundation. 
We thank T. Schneider and M.H.Pedersen for valuable discussions.

\end{document}